\documentclass{epl}

\title{Structure Below the Growing Surface}
\shorttitle{The Structure Below}
\author{E. Katzav \inst{1} \and S. F. Edwards \inst{2} \and M. Schwartz \inst{3}}

\institute{
  \inst{1} Department of Chemical Physics, The Weizmann Institute of Science Rehovot 76100, Israel\\
  \inst{2} Polymers and Colloids Group, Cavendish Laboratory, Cambridge University Madingley Road, CB30HE Cambridge, United Kingdom\\
  \inst{3} School of Physics and Astronomy, Raymond and Beverley Faculty of Exact Sciences, Tel Aviv University, Tel Aviv 69978, Israel
   }


\begin{document}

\maketitle

\begin{abstract}
In recent years there has been a growing interest in the statistical
properties of surfaces growing under deposition of material. Yet it
is clear that a theory describing the evolution of a surface should
at the same time describe the properties of the bulk buried
underneath. Clearly, the structure of the bulk is relevant for many
practical purposes, such as the transport of electric current in
devices, transport of fluids in geological formations and stress
transmission in granular systems. The present paper demonstrates
explicitly how models describing deposition can provide us with
information on the structure of the bulk. Comparison of an analytic
model with a simulation of a discrete growth model reveals an
interesting long range tail in the density-density correlation in
the direction of deposition.
\end{abstract}

%\section{Introduction}
The study of the statistical properties of surfaces growing under
the deposition of material has attracted many researchers over the
last two decades. The systems under consideration vary from heaps of
sand or other assemblies of granular matter, to devices manufactured
by the bombardment of atoms on a growing target. The theoretical
description of such systems is given by a number of discrete and
continuous models that belong mainly to three categories. The first
is the Edwards Wilkinson category \cite{EW} which was constructed to
describe a situation of slow deposition under gravity where each
deposited particle has the time to find its lowest possible
gravitational potential energy in the presence of the existing
surface. The second category is that of KPZ \cite{KPZ} in which
lateral growth is important. This can be a result of sticking or
just the geometry of growth perpendicular to the surface as
explained in ref. \cite{KPZ}. The third category is the MBE
(Molecular Beam Epitaxy) \cite{Wolf90} which was constructed to
describe processes of device fabrication in which the physics should
produce under a wide range of parameters flat surfaces. The focus of
interest in those studies was the statistical characterization of
the growing surface. This is achieved by obtaining the roughness
exponent of the steady state surface, the growth exponent
\cite{Wolf87,KK89,Forrest90,Kim91,SCE92,BC,Halpin95,Barabasi,Perlsman96,SCE98,Edwards02a,Katzav04a,DasSarma91,Lai91,Nattermann91,Kessler92,Katzav02}
and the scaling functions associated with the steady state evolution
of the surface \cite{Edwards02b,Moore01,Moore02,Katzav04b}. It is
easy to envisage many practical applications for which the
fluctuations in the steady state surface are relevant and this was
always one of the motivations for the intensive study of surface
growth. Yet it is obvious that the internal structure of the
material below the surface may be of much more practical importance
as this determines the mechanical properties of the system,
generated by deposition, such as a heap of granular matter in which
it will affect stress transmission. In electrical devices obtained
by deposition, the structure of the bulk will obviously affect the
transmission of electric current which will determine the
functionality of the device. Certain geological formations relevant
to the oil industry are also generated by deposition. The structure
of the bulk determines the important flow properties through the
formation.

This letter has two goals. The first and more important goal is to
draw attention to the fact that a theory which describes the
evolution of the upper surface of a system growing under the
deposition of material, should simultaneously be able to predict the
structure of the bulk below that surface. The physical reason for
our statement is the fact that serious rearrangement does not
usually take place in the deeper layers below the surface.

The second goal is to explicitly demonstrate how the structure of
bulk below the surface can be obtained for two growth models, the
continuous analytical KPZ system and the discrete numerical
ballistic deposition (BD) system. Note, that the main reason for
treating the two models mentioned above, is that both possesses as
will be shown in the following share the common realistic physical
property that the growth process is accompanied by the embedding of
voids below the evolving surface (this property is not shared by
other very successful discrete numerical systems such as SOS
\cite{Meakin86} and RSOS \cite{KK89} which do not allow for the
generation of voids below the surface.) The second reason for
treating these two systems is that they are, widely known and simple
enough to demonstrate the connection between the system describing
the evolution of the surface and the structure of the bulk below it.

\begin{figure}
 \onefigure[width=6cm]{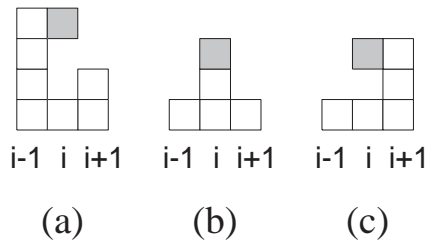}
 \caption{The addition of the shaded
particle creates a void in (a) and (c), where it sticks to the
column on it right or left. No void is created (b).}
 \label{F1}
\end{figure}

To clarify our ideas consider first the Edwards Wilkinson equation
for the local height of the surface $h({\bf{r}},t)$,
\begin{equation}
\frac{{\partial h}}{{dt}} = \nu \nabla ^2 h + R\left( {{\bf{r}},t}
\right)
 \label{E1},
\end{equation}
where $R\left( {{\bf{r}},t} \right)$ is the local rate of deposition
of material given by $R\left( {{\bf{r}},t} \right) = R_0  + \eta
\left( {{\bf{r}},t} \right)$ and $\eta \left( {{\bf{r}},t} \right)$
is a noise term that has zero average and its correlations are
usually taken to be
\begin{equation}
\left\langle {\eta \left( {{\bf{r}},t} \right)\eta \left(
{{\bf{r'}},t'} \right)} \right\rangle  = D\Delta \left( {{\bf{r}} -
{\bf{r'}}} \right)\delta \left( {t - t'} \right)
 \label{E3},
\end{equation}
where $\Delta$ is a short range function. The constant position and
time independent rate $R_0$ is traditionally deleted, because it has
no effect on the shape of the surface. For our purpose though, we
have to keep it. To be precise note that the deposition rate is
related to the local number of particles landing per unit area per
unit time, $n\left( {{\bf{r}},t} \right)$, by the relation $R\left(
{{\bf{r}},t} \right) = \Omega n\left( {{\bf{r}},t} \right)$ where
$\Omega$ is the effective volume taken by each landing particle. Now
particles land and rearrange on the surface to minimize their
potential (gravitational) energy. This rearrangement is described by
the first term on the right hand side of eq. (\ref{E1}). Dividing
$h\left( {{\bf{r}},t} \right)$ by $\Omega$ we obtain the total area
density of particles at the point $\bf{r}$ at time $t$. It is clear
that whatever the dynamical picture of surface evolution is, the
total number of particles has to be conserved and indeed eq.
(\ref{E1}) trivially conserves the number of particles. Actually it
does more than that. It conserves the volume occupied by the
particles. The picture is thus the following. The landing particles
form a compact structure and then rearrange on the surface
preserving the compact structure beneath it. If the particles are
cubes, as in some discrete models, there remain no voids among the
particles. Other shapes of particles must result, of course, in
voids among them but the structure is expected to be either ordered
or random close packing. Our following discussion is not concerned
with that compact structure. We focus rather on deviations from that
structure involving larger voids in the system. The physical reasons
for such voids include sticking and the geometry of growth normal to
the surface, that are described within the KPZ category. This
incorporates voids into the structure in a most natural way. To
visualize it, consider the following version of ballistic
deposition. In this model a particle falls vertically and sticks to
the first site along its trajectory that has an occupied nearest
neighbor. The particle is not allowed to stick to a diagonal
neighbor. The last particle to be deposited is shaded in Fig.
(\ref{F1}) that demonstrates how voids are created. A void, once
created, does not disappear.

In the following we will discuss the problem of structure below the
evolving surface from two different points of view. The first is to
show that within the continuous KPZ description of the evolving
surface, it is possible to obtain the density-density correlation
function of the material below the surface in terms of height
correlations related to the evolution of the surface. The second is
to obtain the density-density correlations directly from a one
dimensional BD simulation.

Consider next the KPZ equation
\begin{equation}
\frac{{\partial h}}{{\partial t}} = \nu \nabla ^2 h + \lambda \left(
{\nabla h} \right)^2  + R({\bf r},t)
 \label{E5}.
\end{equation}
The origin of the non linear term in the above equation is the fact
that growth is perpendicular to the surface. It is clear that if a
particle lands on the surface an outcrop perpendicular to the
surface is generated. The overhang screens the area below it and
prevents more descending particles to fill the void generated by
that particle just as in the discrete BD version depicted in Fig.
(\ref{F1}). The KPZ equation must be thus viewed not as an equation
for a single valued height function, because the height function is
not really single valued, but rather an equation for the height
envelope function below which there is an abundance of voids. Can we
see the existence of those voids from the KPZ equation itself? From
the relation $R({\bf r},t) = \Omega n({\bf r},t)$ it is clear that
the total "deposited volume" is given by $\int {d{\bf{r}}dtR\left(
{{\bf{r}},t} \right)}$ (assuming we start deposition on a flat
surface at time $t=0$). From eq. (\ref{E5}) on the other hand it is
clear that the total volume below the envelop surface is larger than
that by $\lambda \int {d{\bf{r}}dt} \left( {\nabla h\left(
{{\bf{r}},t} \right)} \right)^2$ which must be the volume occupied
by voids. The average of the integrand is known in the literature as
the excess velocity and its meaning is that the average height
growth faster than expected from the rate of descending material and
this is just due to voids being incorporated into the structure
\cite{Barabasi}. Suppose we could identify that part of the local
rate of increase in the height, $\left[ {\frac{{\partial
h({\bf{r}},t)}}{{\partial t}}} \right]_v$, which is a result of void
creation, then the local density $\rho ({\bf{r}},z)$ at a point,
through which the surface passed at time $t$, is given by
\begin{equation}
\rho \left( {{\bf{r}},z} \right) = \rho _0 {{\left\{
{\frac{{\partial h\left( {{\bf{r}},t} \right)}}{{\partial t}} -
\left[ {\frac{{\partial h\left( {{\bf{r}},t} \right)}}{{\partial
t}}} \right]_v } \right\}} \mathord{\left/
 {\vphantom {{\left\{ {\frac{{\partial h\left( {{\bf{r}},t} \right)}}{{\partial t}} - \left[ {\frac{{\partial h\left( {{\bf{r}},t} \right)}}{{\partial t}}} \right]_v } \right\}} {\frac{{\partial h\left( {{\bf{r}},t} \right)}}{{\partial t}}}}} \right.
 \kern-\nulldelimiterspace} {\frac{{\partial h\left( {{\bf{r}},t} \right)}}{{\partial t}}}}
 \label{E6},
\end{equation}
where $\rho_0$ is the constant density that would have existed in
the corresponding Edwards Wilkinson system.

The left hand side of the above gives $\rho$ as a function of $z$
but on the right hand side we have functions of $t$. The next step
must thus be to connect between the perpendicular coordinate $z$ and
the passage time of the surface $t$. This relation is readily
obtained by noting that the height of the surface at the point
${\bf{r}}$ and time $t$ is given by
\begin{equation}
h\left( {{\bf{r}},t} \right) = R_0 t + \delta h\left( {{\bf{r}},t}
\right)
 \label{E7},
\end{equation}
where $\delta h({\bf{r}},t)$ is the usual variable describing the
width of the surface. This quantity can attain values of the order
of a positive power in the lateral size of the system. Nevertheless,
if we wait for long enough times, that make $R_0 t$ ,to be of the
order of the lateral size, the second term on the right hand side of
eq. (\ref{E7}) is negligible compared to the first. The relation can
be thus iterated in the following manner
\begin{equation}
t = {z \mathord{\left/
 {\vphantom {z {R_0 }}} \right.
 \kern-\nulldelimiterspace} {R_0 }} - \delta h\left( {{\bf{r}},{z \mathord{\left/
 {\vphantom {z {R_0 }}} \right.
 \kern-\nulldelimiterspace} {R_0 }} - \delta h({\bf{r}},z/R_0 \dots)} \right)
 \label{E8}.
\end{equation}
Within the leading approximation to eq. (\ref{E8}) eq. (\ref{E6}) is
also simplified and reads
\begin{equation}
\rho \left( {{\bf{r}},z} \right) = \rho _0 \left\{ {1 - {{\left[
{\frac{{\partial h\left( {{\bf{r}},t} \right)}}{{\partial t}}}
\right]_{v,t = {z \mathord{\left/
 {\vphantom {z {R_0 }}} \right.
 \kern-\nulldelimiterspace} {R_0 }}} } \mathord{\left/
 {\vphantom {{\left[ {\frac{{\partial h\left( {{\bf{r}},t} \right)}}{{\partial t}}} \right]_{v,t = {z \mathord{\left/
 {\vphantom {z {R_0 }}} \right.
 \kern-\nulldelimiterspace} {R_0 }}} } {{\rm{R}}_{\rm{0}} }}} \right.
 \kern-\nulldelimiterspace} {{\rm{R}}_{\rm{0}} }}} \right\}
 \label{E9}.
\end{equation}
We identify next
\begin{equation}
\left[ {\frac{{\partial h\left( {{\bf{r}},t} \right)}}{{\partial
t}}} \right]_v  = \lambda \left( {\nabla h\left( {{\bf{r}},t}
\right)} \right)^2
 \label{E10},
\end{equation}
since this non-linear term is the term responsible for the "excess
velocity" and thus for the incorporation of voids. Eqs. (\ref{E9})
and (\ref{E10}) will be used now to obtain the average density $\bar
\rho$ and the density-density correlations in the system
\begin{equation}
\bar \rho  = \rho _0 \left( {1 - {{\lambda \left\langle {\left(
{\nabla h} \right)^2 } \right\rangle } \mathord{\left/
 {\vphantom {{\lambda \left\langle {\left( {\nabla h} \right)^2 } \right\rangle } {R_0 }}} \right.
 \kern-\nulldelimiterspace} {R_0 }}} \right)
 \label{E11}.
\end{equation}
Now,assuming that the steady state structure factor is given by
$\left\langle {h_{\bf{q}} h_{ - {\bf{q}}} } \right\rangle =
Aq^{-\Gamma}$ for $q \le q_0$ (where $q_0$ is the high momentum
cut-off, corresponding to the size of landing particles) and zero
above it, the average density is given by
\begin{equation}
\bar \rho  = \rho _0 \left[ {1 - {{\lambda AS_d q_0^{(d + 1 - \Gamma
)} } \mathord{\left/
 {\vphantom {{\lambda AS_d q_0^{(d + 1 - \Gamma )} } {\left( {d + 1 - \Gamma } \right)R_0 }}} \right.
 \kern-\nulldelimiterspace} {\left( {d + 1 - \Gamma } \right)R_0 }}} \right]
 \label{E12},
\end{equation}
where $d+1$ is the dimension of space and $S_d$ is the surface area
of a unit sphere in $d$ dimensions. The density-density correlations
involve one non-trivial correlation
\begin{equation}
\Psi \left( {{\bf{r}},t} \right) = \left\langle {\left[ {\nabla
h({\bf{r}},t)} \right]^2 \left[ {\nabla h(0,0)} \right]^2 }
\right\rangle  - \left[ {\left\langle {\left( {\nabla h} \right)^2 }
\right\rangle } \right]^2
 \label{E13}.
\end{equation}
To evaluate it we calculate it to the lowest order in the frequency
dependent structure factor $\Phi ({\bf{q}},\omega ) = \left\langle
{h_{{\bf{q}},\omega } h_{ - {\bf{q}}, - \omega } } \right\rangle$.
Standard manipulation leads to the expression
\begin{equation}
\Psi \left( {{\bf{r}},t} \right) =  - \frac{2}{{\left( {2\pi }
\right)^{2d} }}A^2 \int {d{\bf{l}}d{\bf{l'}}\left( {{\bf{l}} \cdot
{\bf{l'}}} \right)^2 l^{ - \Gamma } l'^{ - \Gamma } f\left( {\omega
_l t} \right)f\left( {\omega _{l'} t} \right)\exp \left[ {i\left(
{{\bf{l}} + {\bf{l'}}} \right) \cdot {\bf{r}}} \right]}
 \label{E14},
\end{equation}
where $\omega_l = Bl^z$ is the typical frequency related to the
decay of a disturbance of wave vector $\bf{l}$ and the scaling
function $f(u)$ is related to the time dependent structure factor
$\Phi \left( {{\bf{q}},t} \right) = \left( {2\pi } \right)^{ - 1}
\int {d\omega \Phi \left( {{\bf{q}},\omega } \right)e^{i\omega t}}$
by
\begin{equation}
\Phi \left( {{\bf{q}},t} \right) = Aq^{ - \Gamma } f\left( {\omega
_q t} \right)
 \label{E15}.
\end{equation}

Clearly, the calculation of higher orders in the iteration may be
rather tedious but they are straightforward as described in refs.
\cite{SCE98,Edwards02a}. From our point of view, however, the
important statement is that density-density correlations below the
growing surface are related to various height correlations which are
natural objects of study in the traditional research of surface
growth. It has to be carried in mind, though, that to use the
correlation $\Psi \left( {{\bf{r}},t} \right)$ for our purpose by
relating the time difference to the difference in height, we must
have $R_0 t \gg \left| {\delta h\left( {{\bf{r}},0} \right)}
\right|,\;\left| {\delta {\rm{h}}\left( {{\rm{0}}{\rm{,t}}} \right)}
\right|$.

We will work out as an example the one dimensional case and compare
it with the simulation of the one dimensional ballistic deposition.
The density-density correlation function $g\left( {x,y} \right) =
\left\langle {\rho \left( x \right)\rho \left( y \right)}
\right\rangle  - \bar \rho ^2$ is related to $\Psi$ in the one
dimensional case by
\begin{equation}
g\left( {x,y} \right) = \left( {\frac{\lambda }{{{\rm{R}}_{\rm{0}}
}}} \right)^2 \rho _0^2 \Psi \left( {x,{\textstyle{y \over {R_0 }}}}
\right)
 \label{E16}.
\end{equation}
Using the scaling form for the 1D time dependent structure factor
\cite{Barabasi}, $\Phi \left( {q,t} \right) = \frac{D}{\nu }q^{ - 2}
f\left( {Bq^{{3 \mathord{\left/
 {\vphantom {3 2}} \right. \kern-\nulldelimiterspace} 2}} t}\right)$
(where $D$ is the noise amplitude, $\nu$ the diffusion coefficient
and $B$ a numerical constant which cannot be determined
analytically) we obtain
\begin{equation}
\Psi \left( {x,t} \right) =  - \frac{{8D}}{{9\pi ^2 \nu ^2 B^{{4
\mathord{\left/
 {\vphantom {4 3}} \right.
 \kern-\nulldelimiterspace} 3}} }}\frac{1}{{t^{{4 \mathord{\left/
 {\vphantom {4 3}} \right.
 \kern-\nulldelimiterspace} 3}} }}\left[ {\int\limits_0^\infty  {\frac{1}{{\tau ^{{1 \mathord{\left/
 {\vphantom {1 3}} \right.
 \kern-\nulldelimiterspace} 3}} }}f\left( \tau  \right)\cos \left\{ {\left( {\frac{\tau }{{Bt}}} \right)^{{2 \mathord{\left/
 {\vphantom {2 3}} \right.
 \kern-\nulldelimiterspace} 3}} x} \right\}d\tau } } \right]^2
 \label{E18}.
\end{equation}
This form is based on the scaling form and is therefore correct for
large $t$, so that for large $y$ we obtain a power law behavior for
\begin{equation}
g\left( {0,y} \right) =  - \left( {\frac{\lambda
}{{{\rm{R}}_{\rm{0}} }}} \right)^2 \rho _0^2 \frac{{8D}}{{9\pi ^2
\nu ^2 B^{{4 \mathord{\left/
 {\vphantom {4 3}} \right.
 \kern-\nulldelimiterspace} 3}} }}\frac{1}{{\left( {{y \mathord{\left/
 {\vphantom {y {R_0 }}} \right.
 \kern-\nulldelimiterspace} {R_0 }}} \right)^{{4 \mathord{\left/
 {\vphantom {4 3}} \right.
 \kern-\nulldelimiterspace} 3}} }}\left[ {\int\limits_0^\infty  {\frac{1}{{\tau ^{{1 \mathord{\left/
 {\vphantom {1 3}} \right.
 \kern-\nulldelimiterspace} 3}} }}f\left( \tau  \right)d\tau } } \right]^2  \propto  - y^{ - {4 \mathord{\left/
 {\vphantom {4 3}} \right.
 \kern-\nulldelimiterspace} 3}}
 \label{E19}.
\end{equation}

It is interesting now to discuss briefly the case $\lambda < 0$. The
parameters $R_0$ and $\lambda$ corresponding to our physical picture
are positive, of course. The situation with positive $R_0$ and
negative $\lambda$, which can be handled formally by our equations,
has no easy natural interpretation. Certainly this is not related to
the RSOS models, which are supposed to correspond to negative
$\lambda$ but produce compact structures.

Let us turn now to the density-density correlations obtained from a
one dimensional model of ballistic deposition. The model used is the
nearest-neighbor ballistic deposition (NNBD) model, on a lattice
with $L=1024$. At each time step a column $i$ is picked at random. A
particle (just one particle!) falls vertically and sticks to the
first site along its trajectory that has an occupied nearest
neighbor as shown in Fig. \ref{F1}. We found it simpler not to use
periodic boundary conditions. Fig. \ref{F2} below presents the
structure obtained after deposition. The alternating shades of grey
correspond to particles deposited within different time intervals.
The voids are the white regions. This figure corresponds to a
lateral size of $512$ sites.
\begin{figure}
 \onefigure[height=5cm,width=7cm]{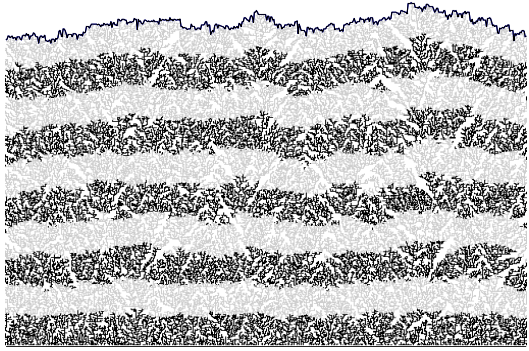}
 \caption{A BD cluster obtained by depositing $100,000$ particles on a substrate of size $L=512$. A time step is defined by a deposition of a single particle. The different shadings correspond to different time intervals each corresponding to the deposition of $10,000$ particles.}
 \label{F2}
\end{figure}

\begin{figure}
 \twoimages[width=5.5cm]{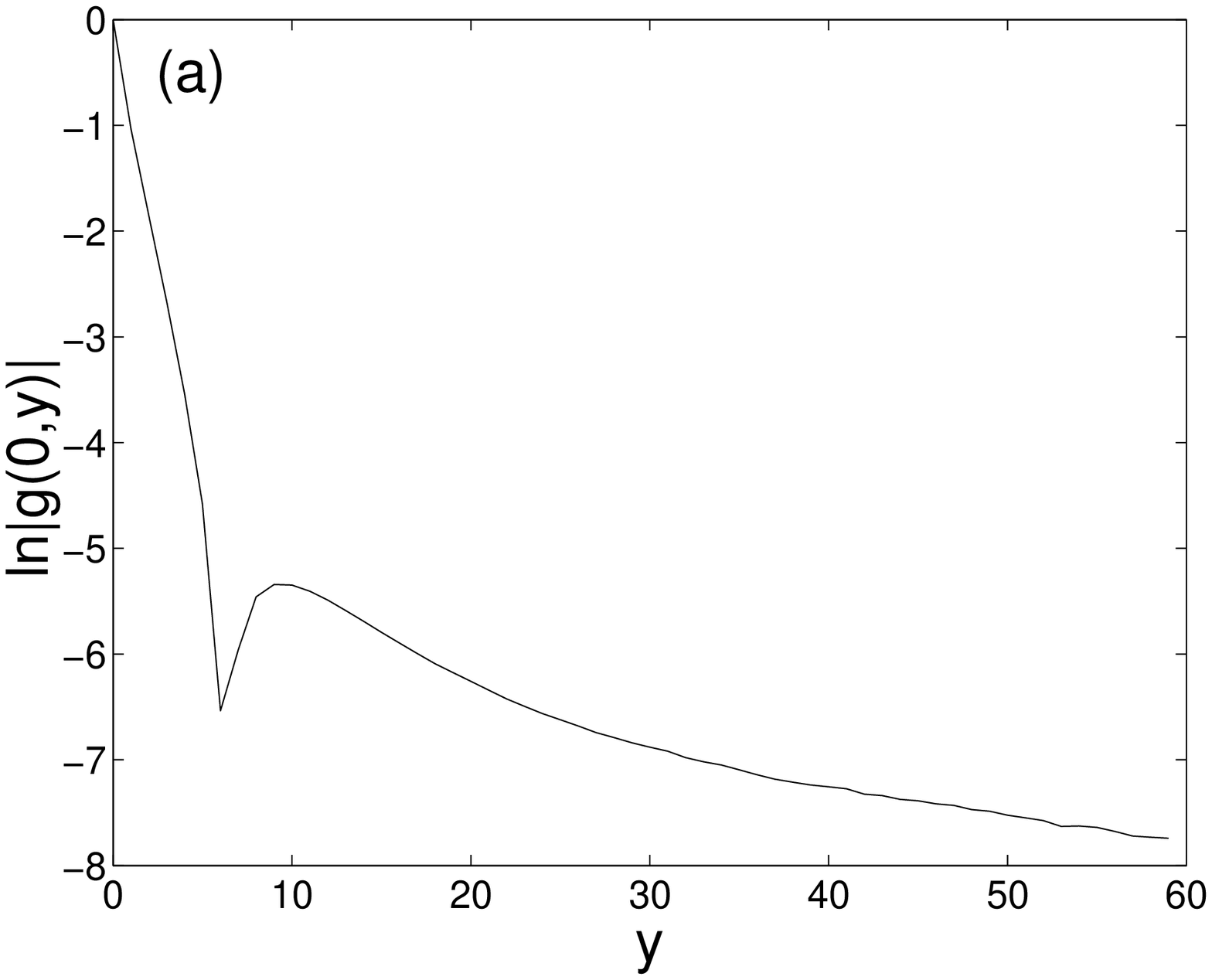}{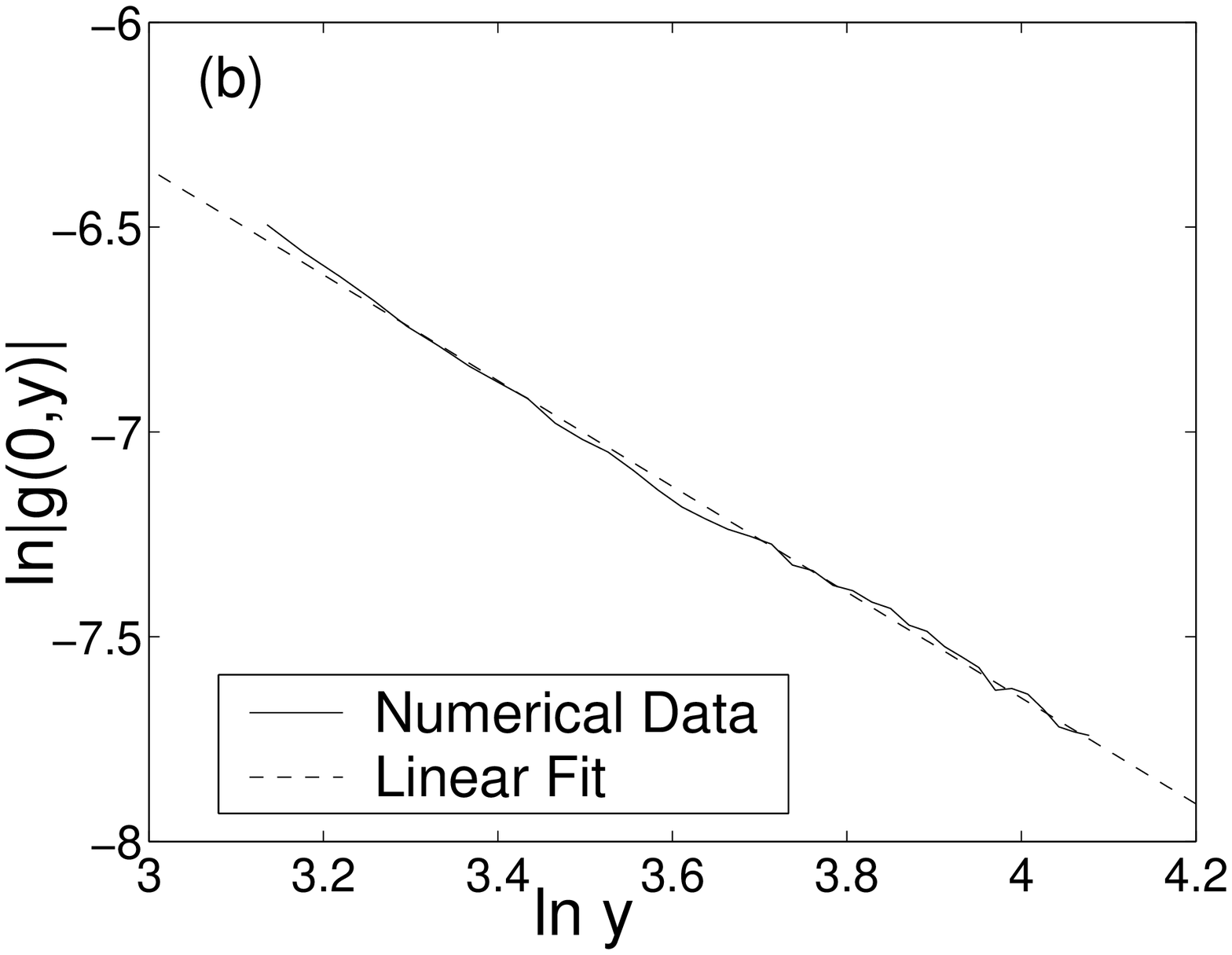}
 \caption{(a) The log of the correlation function in the growth direction. (b) A log-log plot of the tail.}
 \label{F3}
\end{figure}

The long-range correlations obtained in the analytical calculation
above can be already observed qualitatively in the plot of the
cluster generated by the simulation (Fig. \ref{F2} below), where
some the voids tend to extend upward for a long distance. In order
to quantify this, we obtain the correlation $g(0,y)$ from the
simulation. We define the correlation function
\begin{equation}
g\left( {x,y} \right) = \frac{1}{N}\sum {\left\langle {\rho \left(
{\bf{r}} \right)\rho \left( {{\bf{r'}}} \right)} \right\rangle }  -
\bar \rho ^2
 \label{E20},
\end{equation}
where $x$ is in the lateral direction, $y$ is in the perpendicular
(growth) direction, ${\bf{r}} = \left( {X,Y} \right)$, ${\bf{r'}} =
\left( {X',Y'} \right)$ such that $\left| {X - X'} \right| = x$ and
$\left| {Y - Y'} \right| = y$. The number of such pairs $\left(
{{\bf{r}},{\bf{r'}}} \right)$ is $N$, the average is over
realizations of the randomness and $\bar \rho$ is the average
density. Practically we take the average to be the average over
runs. In this work we focused on correlations in the $y$ (growth)
direction since they exhibit nontrivial features such as algebraic
tails. We depict $g(0,y)$ in Fig. \ref{F3}. The total number of
pairs that goes into the evaluation of $g(0,y)$, which is the
product of the number of pairs in one run times the number of runs
is  $2.4 \times 10^{10}$.

We depict the log of the absolute value correlation function,
because the correlation function drops fast such that it is
difficult to observe certain features. For example, it is difficult
to see that the correlations become negative. When we depict the
logarithm of the absolute value we see the point where the function
changes sign as a kink. (Because the $y$-axis is a discrete
coordinate the correlations do not become zero but go from positive
values where they are decreasing to negative value where they are
increasing). The correlation drops very fast and may seem to be of
short range. The analytical prediction though suggests that we
should examine more carefully the tail of $g(0,y)$. In Fig.
\ref{F3}(b) we depict a log-log plot of this tail, which reveals a
power-law behavior or the tail, namely $g\left( {0,y} \right)
\propto -y^{-1.301}$.  We see good agreement the between the
numerical result given above and the analytical one. Both yield a
negative tail decreasing in size as a power law with exponents that
are quite close ($1.3$ vs. $1.333$).

We have chosen here to concentrate on the density-density
correlation function, because it is the most obvious and traditional
characterization of the structure. It is clear that for specific
applications like stress transmission or transmission of electrical
currents we might be interested in other attributes of the
structure. The main point we make in this letter is that the same
analytical and numerical techniques used for the study of the growth
of surfaces can be used in the study of material structure below the
surface. Once this point is realized other properties of the
structure can be readily be obtained. In addition, we have obtained
correlations in the growth direction with an unexpected algebraic
tail. If this is not just an artifact of the two models studied, it
may have far reaching practical and theoretical consequences. As a
possible usage we claim that the anisotropy between the growth
direction and the lateral growth is a signature of the growth
direction. Thus, having at hand a specimen of a sedimentary rock,
for example, our result suggest that one can determine its original
growth direction by determining the principal direction at which an
algebraic tail in the structure factor appears.

We hope that the present work will trigger more research in this
direction, both experimentally, and by introduction of more
realistic discrete models of deposition.

\acknowledgments
The research of E. Katzav at the Weizmann institute
of Science has been supported by the Edith and Edward F. Anixter
postdoctoral fellowship.

\end{document}